# Directional field-dependence of tunable magnetic domains in noncentrosymmetric ferromagnetic Weyl semimetal CeAlSi

Bochao Xu[1], Jacob Franklin[1], Hung-Yu Yang[2], Fazel Tafti[2] and Ilya Sochnikov[1,3,*]

[1]*Physics Department, University of Connecticut, Storrs, CT USA, 06269*

[2]*Department of Physics, Boston College, Chestnut Hill, MA 02467, USA*

[3]*Institute of Material Science, University of Connecticut, Storrs, CT USA, 06269*

[*]*Corresponding author: ilya.sochnikov@uconn.edu*

Dynamics and textures of magnetic domain walls (DWs) may largely alter the electronic behaviors in a Weyl semimetal system via emergent gauge fields. However, very little is known about even the basic properties of these domain walls in Weyl materials. In this work, we imaged the spontaneous magnetization and magnetic susceptibility of a ferromagnetic (FM) Weyl semimetal CeAlSi using scanning SQUID microscopy. We observed the ferromagnetic DWs lined-up with the [100] direction (or other degenerate directions). We also discovered the coexistence of stable and metastable domain phases, which arise likely due to magnetoelastic and magnetostriction effects and are expected to be highly tunable with small strains. We applied an in-plane external field as the CeAlSi sample was cooled down to below the magnetic phase transition of 8.3K, showing that the pattern of FM domains is strongly correlated with both the amplitude and the orientation of the external field even for weak fields of a few Gauss. The area of stable domains increases with field and reaches maximum when the field is parallel to the main crystallographic axes of the CeAlSi crystal. Our results suggest that the manipulation of these heterogeneous phases can provide a practical way to study the interplay between magnetism and electronic properties in Weyl systems, and that these systems can even serve as a new platform for magnetic sensors.

Introduction

The experimental realization of Weyl semimetals is one of the most exciting developments in condensed matter physics in the past decade [1]. Weyl fermions, namely the massless spin-1/2 quasiparticles around band crossing points, were described by the Weyl equations derived by Hermann Weyl in 1929. These band touching points, protected by breaking either spatial-inversion symmetry or time reversal symmetry, are called Weyl points. The first observed Weyl semimetal was non-magnetic TaAs with broken spatial-inversion symmetry [2,3]. Due to the magnetic disorder brought by domains and other intrinsic magnetic structures, magnetic Weyl semimetals with broken time reversal symmetry were not observed until four years later. In 2019, angle-resolved photoemission spectroscopy (ARPES) and scanning tunneling microscopy (STM) studies revealed a linear dispersion and Fermi arcs in Kagome-lattice $Co_3Sn_2S_2$ [4–6], and hence lead to the discoveries of many fascinating phenomena, such as the anomalous Hall effect (AHE) and Nernst effect [7].

Magnetic texture, i.e. magnetic domains, in the Weyl semimetals plays a significant role in its unique electronic behavior [8]. Domains may induce localized charges on the domain walls and therefore can be regarded as a source of axial gauge fields [9–11]. Transport properties of the Weyl semimetals are expected to be tunable with these axial gauge fields [9], [12,13]. Overall, manipulation of magnetic domains in Weyl semimetals is a promising avenue for topologic electronic engineering, which is what motivated this work.

CeAlSi is a non-centrosymmetric ferromagnetic Weyl semimetal with strong magnetic anisotropy [14,15]. It shows both AHE and loop Hall effect corresponding to an in-plane or out-of-plane external magnetic field [15]. Our previous study revealed a coexistent state of stable and metastable domains inside CeAlSi. The domain structure is sensitive to external fields and temperature changes via a metastable phase [16]. Previously, we measured sample's magnetoelasticity by a strain-gauge approach, showing that small strains, less than one picometer per unit cell, can be induced by magnetoelastic effect [16]. The magnetoelastic effect likely originates in a simple torque interaction between the Ce moments and the internal magnetic field  The strain was responsible for formation of the field-dependent metastable states [16] . In this work, we further study how the field strength and orientation may alter the distribution of stable domains in CeAlSi.

Experimental

Scanning SQUID experiments were done at the University of Connecticut using a Montana instruments Fusion 2 Cryostation, with a home-built microscope [17].

Scanning SQUID microscopy can resolve micrometer-scale variations in both magnetic and susceptibility signatures in the material under study. The SQUID used in this work is a gradiometric susceptometer, which can be thought of as a miniaturized version of the mutual-inductance setup in the reflection geometry [18,19]. The dimensions of our SQUID device is 7 and 3.25 $\mu m$ for the field coil and pickup loop mean radii, respectively. Scanning the SQUID parallel to the film surface thus allows for micrometer-scale imaging. A small local AC field from the SQUID field coil induces a response from the sample, and the SQUID measures that response via the pick-up loop (Figure 1 (a)). Gradiometric modulation coils positioned at the center of the device couple to the response, and we measure the modulation current in well-calibrated flux units via a feedback loop amplifier. The SQUID sensor used in this work can, in principle, operate with few hundred Gauss in-plane external field. However, it is more stable and less noisy at lower fields. Therefore, in this work we used only tens of Gausses to study the angular field-dependence of the domains' texture. In order to collect images, a piezo-scanner moves in a raster fashion to record the magnetic susceptibility and the dc flux distribution in a rectangular scan area [20–22,15,16].

CeAlSi crystals were grown using a self-flux method [15]. The starting materials Ce, Al, and Si were mixed in a mole ratio of 1:10:1 and placed in an alumina crucible inside an evacuated silica tube. The mixture was heated to 1000 °C for 12 h, cooled to 700 °C at 0.1 °C/min, annealed at 700 °C for 12 hours and then centrifuged to decant the residual Al-flux. The crystals can be up to several milliliters wide with a metallic luster, and they are stable in air. The CeAlSi samples have a ferromagnetic transition temperature of 8.3 K. We polished the (001) surface of the samples using a standard polishing jig and a fine diamond lapping film to obtain a flat surface suitable for the scanning SQUID imaging. An $I4_1md$ space group structure and the ferromagnetic phase transition were previously verified by optical second harmonic generation and neutron scattering [15].

Results and Discussion

In this work, we studied a crystal that was larger than previously [15,16], with a size of about 1.7 x1 mm$^2$), which showed less of the metastable regions [16], possibly due to a size-effect on the magnetostriction strains, and thus mostly stable domains with narrow DWs were observed (Figure 1 (b)-(c)). Magnetic domains were measured by applying a small field normal to the *c*-axis of a CeAlSI crystals (Figure 1 d).

Figure 2 shows the domains' texture in a field-cooled CeAlSi crystal over a range of field orientation. External fields were applied above the transition temperature of 8.3 K. Scanning images were taken after the sample was cooled down to 5.3 K. As shown in both in-phase susceptibility and magnetization channels, the orientation of a small external field could effectively alter the landscape of the domains. When the field is along the crystallographic [100] direction (~67.5 degree in our frame of reference), a large fraction of the domains are in a tall narrow rectangular shape with long sides lined-up with the [010] direction, normal to the external field. These long narrow domains expand in the [100] direction and shrink in the [010] direction as the field was rotated to other discrete angles counterclockwise. The number of domains decreases and large domains in non-uniform shape are developed due to the collapse of the domain walls in the [010] direction when the field is off the main crystal axis, as illustrated in Figure 1 (c).

Figure 3 quantitatively shows how domain wall lengths distribute in both the [100] and [010] direction. Domain walls along [010] are dominant when the field is oriented to 45°, 67.5°, 90° and 112.5°. The projection of the field contributes more to [100] than [010] at these angles. The lengths of the [010] domain walls reach a maximum at 67.5° and then drop as the field rotates away from the [100] direction. At 135°, where the field is more than 45° to the [100] direction, the domain walls pointing to [100] exhibit a sharp increase in length and outnumber the domain walls along the [010] direction. The unexpected raise in domain wall lengths at 45° is suppressed when the magnitude of the external field is increased from 13.2G to 22G, which suggests that stronger magnetic fields may have better performance in the manipulation of domain distribution. The insets to Figure 3 emphasize how the transition of the domain orientation happens near the field orientated along [110] direction; the sharpness of this transition and its influence on the anisotropic resistance is a subject of our future investigations in of CeAlSi and other related materials.

Summary and Conclusions

In summary, we have presented magnetic imaging results for the noncentrosymmetric ferromagnetic Weyl semimetal CeAlSi that show how magnetic domains and domain walls can be controlled by tuning the direction of a very small external field. We showed that the field orientation can increase domain wall lengths, with a median increase of 165 $\mu m$, and maximal increase within the field of view of 410 $\mu m$. Alternatively, the area of the domains can be maximized to more than an area of about 200 x 200 $\mu m^2$, when the density of the domains is minimized.

It remains to be studied in future works how these configurations depend on the sample size and their shape anisotropy. One of the main goals of the research into magnetic Weyl semimetals such as RAlX [14] is to correlate highly anisotropic resistances [23–25,15] and incommensurate behavior [8] with the well-controlled domain configurations for use in well-defined devices. This can lead to the confirmation of the emergent gauge fields in these materials and their use for magnetic field sensing and new types of spintronic devices. Our work is a step towards this goal. We believe that simultaneous magnetic imaging and electronic (and spin) transport experiments are the key to success in realizing these new types of devices, particularly quantum sensors made from topologically non-trivial materials.

Acknowledgments

We thank J. R. Kirtley for the inductance calculation software code, the State of Connecticut for financing construction of the scanning SQUID, the College for Liberal Arts and Sciences of the University of Connecticut for special graduate research assistantship support for B. X. and J. F., I. S. acknowledges the US DOD for partial support. The work at Boston College was funded by the National Science Foundation under grant no. DMR-1708929.

References

[1] B. Yan and C. Felser, *Topological Materials: Weyl Semimetals*, Annu. Rev. Condens. Matter Phys. **8**, 337 (2017).
[2] B. Q. Lv, H. M. Weng, B. B. Fu, X. P. Wang, H. Miao, J. Ma, P. Richard, X. C. Huang, L. X. Zhao, G. F. Chen, Z. Fang, X. Dai, T. Qian, and H. Ding, *Experimental Discovery of Weyl Semimetal TaAs*, Phys. Rev. X **5**, 031013 (2015).

[3] S.-Y. Xu, I. Belopolski, N. Alidoust, M. Neupane, G. Bian, C. Zhang, R. Sankar, G. Chang, Z. Yuan, C.-C. Lee, S.-M. Huang, H. Zheng, J. Ma, D. S. Sanchez, B. Wang, A. Bansil, F. Chou, P. P. Shibayev, H. Lin, S. Jia, and M. Z. Hasan, *Discovery of a Weyl Fermion Semimetal and Topological Fermi Arcs*, Science **349**, 613 (2015).

[4] N. Morali, R. Batabyal, P. K. Nag, E. Liu, Q. Xu, Y. Sun, B. Yan, C. Felser, N. Avraham, and H. Beidenkopf, *Fermi-Arc Diversity on Surface Terminations of the Magnetic Weyl Semimetal $Co_3Sn_2S_2$*, Science **365**, 1286 (2019).

[5] D. F. Liu, A. J. Liang, E. K. Liu, Q. N. Xu, Y. W. Li, C. Chen, D. Pei, W. J. Shi, S. K. Mo, P. Dudin, T. Kim, C. Cacho, G. Li, Y. Sun, L. X. Yang, Z. K. Liu, S. S. P. Parkin, C. Felser, and Y. L. Chen, *Magnetic Weyl Semimetal Phase in a Kagomé Crystal*, Science **365**, 1282 (2019).

[6] A. Noah, F. Toric, T. D. Feld, G. Zissman, A. Gutfreund, D. Tsruya, T. R. Devidas, H. Alpern, H. Steinberg, M. E. Huber, J. G. Analytis, S. Gazit, E. Lachman, and Y. Anahory, *Hidden Spin-Texture at Topological Domain Walls Drive Exchange Bias in a Weyl Semimetal*, ArXiv:2101.11639 Cond-Mat (2021).

[7] E. Liu, Y. Sun, N. Kumar, L. Muechler, A. Sun, L. Jiao, S.-Y. Yang, D. Liu, A. Liang, Q. Xu, J. Kroder, V. Süß, H. Borrmann, C. Shekhar, Z. Wang, C. Xi, W. Wang, W. Schnelle, S. Wirth, Y. Chen, S. T. B. Goennenwein, and C. Felser, *Giant Anomalous Hall Effect in a Ferromagnetic Kagome-Lattice Semimetal*, Nat. Phys. **14**, 1125 (2018).

[8] J. Gaudet, H.-Y. Yang, S. Baidya, B. Lu, G. Xu, Y. Zhao, J. Rodriguez, C. M. Hoffman, D. E. Graf, D. H. Torchinsky, P. Nikolic, D. Vanderbilt, F. Tafti, and C. L. Broholm, *Incommensurate Magnetism Mediated by Weyl Fermions in NdAlSi*, ArXiv:2012.12970 (2021).

[9] D. Destraz, L. Das, S. S. Tsirkin, Y. Xu, T. Neupert, J. Chang, A. Schilling, A. G. Grushin, J. Kohlbrecher, L. Keller, P. Puphal, E. Pomjakushina, and J. S. White, *Magnetism and Anomalous Transport in the Weyl Semimetal PrAlGe: Possible Route to Axial Gauge Fields*, Npj Quantum Mater. **5**, 5 (2020).

[10] Y. Araki, A. Yoshida, and K. Nomura, *Localized Charge in Various Configurations of Magnetic Domain Wall in a Weyl Semimetal*, Phys Rev B **98**, 045302 (2018).

[11] Y. Araki, A. Yoshida, and K. Nomura, *Universal Charge and Current on Magnetic Domain Walls in Weyl Semimetals*, Phys Rev B **94**, 115312 (2016).

[12] S. Howlader, R. Ramachandran, Y. Singh, and G. Sheet, *Domain Structure Evolution in the Ferromagnetic Kagome-Lattice Weyl Semimetal $Co_3Sn_2S_2$*, J. Phys. Condens. Matter **33**, 075801 (2020).

[13] J. D. Hannukainen, A. Cortijo, J. H. Bardarson, and Y. Ferreiros, *Electric Manipulation of Domain Walls in Magnetic Weyl Semimetals via the Axial Anomaly*, ArXiv:2012.12785 Cond-Mat (2020).

[14] G. Chang, B. Singh, S.-Y. Xu, G. Bian, S.-M. Huang, C.-H. Hsu, I. Belopolski, N. Alidoust, D. S. Sanchez, H. Zheng, H. Lu, X. Zhang, Y. Bian, T.-R. Chang, H.-T. Jeng, A. Bansil, H. Hsu, S. Jia, T. Neupert, H. Lin, and M. Z. Hasan, *Magnetic and Noncentrosymmetric Weyl Fermion Semimetals in the RAlGe Family of Compounds (R=rare Earth)*, Phys. Rev. B **97**, 041104 (2018).

[15] H.-Y. Yang, B. Singh, J. Gaudet, B. Lu, C.-Y. Huang, W.-C. Chiu, S.-M. Huang, B. Wang, F. Bahrami, B. Xu, J. Franklin, I. Sochnikov, D. E. Graf, G. Xu, Y. Zhao, C. M. Hoffman, H. Lin, D. H. Torchinsky, C. L. Broholm, A. Bansil, and F. Tafti, *A New Noncollinear Ferromagnetic Weyl Semimetal with Anisotropic Anomalous Hall Effect*, Phys. Rev. B **103**, 115143 (2021).

[16] B. Xu, J. Franklin, A. Jayacody, H.-Y. Yang, F. Tafti, and I. Sochnikov, *Back Cover: Picoscale Magnetoelasticity Governs Heterogeneous Magnetic Domains in a Noncentrosymmetric Ferromagnetic Weyl Semimetal (Adv. Quantum Technol. 3/2021)*, Adv. Quantum Technol. **4**, 2170033 (2021).

[17] C. Herrera, J. Franklin, I. Božović, X. He, and I. Sochnikov, *Scanning SQUID Characterization of Extremely Overdoped $La_{2-x}Sr_xCuO_4$*, Phys. Rev. B **103**, 024528 (2021).

[18] M. E. Huber, N. C. Koshnick, H. Bluhm, L. J. Archuleta, T. Azua, P. G. Björnsson, B. W. Gardner, S. T. Halloran, E. A. Lucero, and K. A. Moler, *Gradiometric Micro-SQUID Susceptometer for Scanning Measurements of Mesoscopic Samples*, Rev. Sci. Instrum. **79**, 053704 (2008).

[19] J. R. Kirtley, B. Kalisky, J. A. Bert, C. Bell, M. Kim, Y. Hikita, H. Y. Hwang, J. H. Ngai, Y. Segal, F. J. Walker, C. H. Ahn, and K. A. Moler, *Scanning SQUID Susceptometry of a Paramagnetic Superconductor*, Phys. Rev. B **85**, 224518 (2012).

[20] B. Kalisky, J. R. Kirtley, J. G. Analytis, J.-H. Chu, A. Vailionis, I. R. Fisher, and K. A. Moler, *Stripes of Increased Diamagnetic Susceptibility in Underdoped Superconducting $Ba(Fe_{1-x}Co_x)_2As_2$ Single Crystals: Evidence for an Enhanced Superfluid Density at Twin Boundaries*, Phys. Rev. B **81**, 184513 (2010).

[21] I. Sochnikov, A. J. Bestwick, J. R. Williams, T. M. Lippman, I. R. Fisher, D. Goldhaber-Gordon, J. R. Kirtley, and K. A. Moler, *Direct Measurement of*

Current-Phase Relations in Superconductor/Topological Insulator/Superconductor Junctions, Nano Lett. **13**, 3086 (2013).

[22] I. Sochnikov, L. Maier, C. A. Watson, J. R. Kirtley, C. Gould, G. Tkachov, E. M. Hankiewicz, C. Brüne, H. Buhmann, L. W. Molenkamp, and K. A. Moler, *Nonsinusoidal Current-Phase Relationship in Josephson Junctions from the 3D Topological Insulator HgTe*, Phys. Rev. Lett. **114**, 066801 (2015).

[23] H. Hodovanets, C. J. Eckberg, P. Y. Zavalij, H. Kim, W.-C. Lin, M. Zic, D. J. Campbell, J. S. Higgins, and J. Paglione, *Single-Crystal Investigation of the Proposed Type-II Weyl Semimetal CeAlGe*, Phys. Rev. B **98**, 245132 (2018).

[24] T. Suzuki, L. Savary, J.-P. Liu, J. W. Lynn, L. Balents, and J. G. Checkelsky, *Singular Angular Magnetoresistance in a Magnetic Nodal Semimetal*, Science **365**, 377 (2019).

[25] H. Hodovanets, C. J. Eckberg, Y. Eo, D. J. Campbell, P. Y. Zavalij, P. Piccoli, T. Metz, H. Kim, J. S. Higgins, and J. Paglione, *Anomalous Symmetry Breaking in Weyl Semimetal CeAlGe*, ArXiv:2101.10411 Cond-Mat (2021).

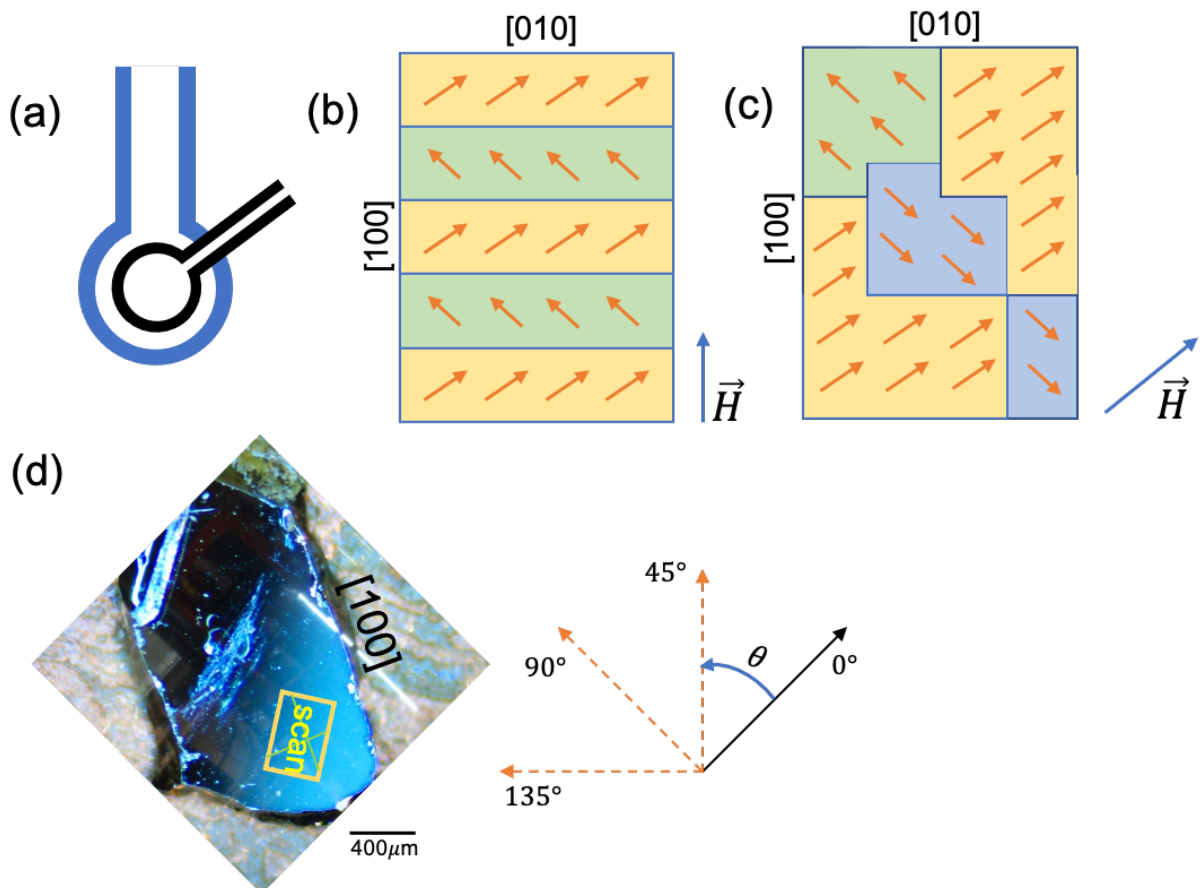

*Figure 1. Main experimental details of scanning SQUID imaging and interpretation of the in-plane domain texture presented in the text. (a) is a schematic of the SQUID sensor used in this work. The SQUID sensor consists of a pickup loop (black) with an*

inner diameter of 3μm and an outer diameter of 3.5μm and a field coil (blue) with an inner diameter of 6μm and an outer diameter of 7μm. (b) schematically shows our interpretation of the in-plane domain texture for two different field orientations. For [100] fields the domain wall size is maximized; for [110] fields the domain area is maximized. The left figure in (c) is the top view of the CeAlSi crystal imaged in this work. The yellow rectangle represents the scanning area. The green lines inside indicate the predominant orientation of the domain walls. The right figure in (c) shows the orientation of the external magnetic field. The orange arrows correspond to fields oriented at 45°, 90° and 135°.

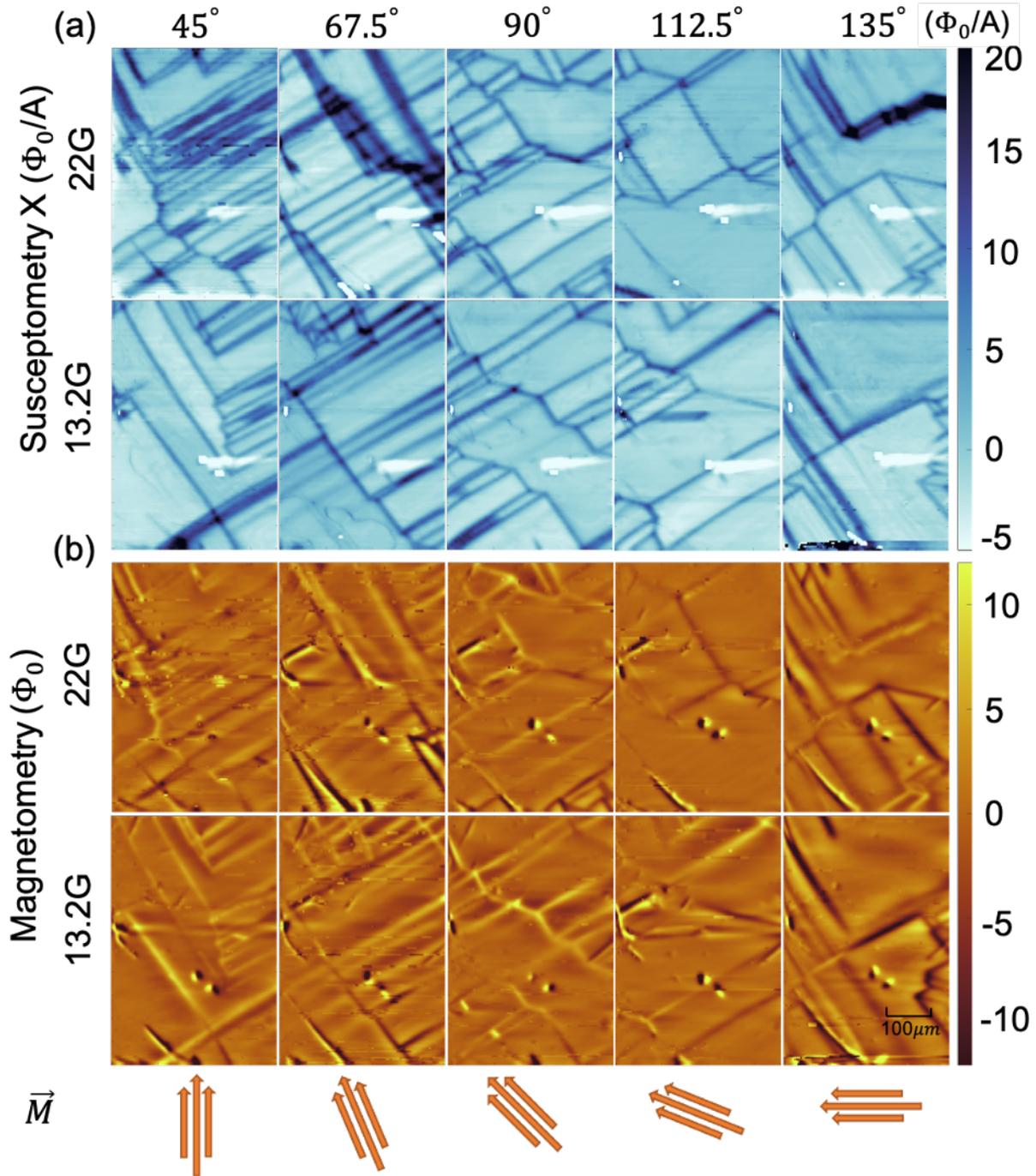

Figure 2. The field orientation dependence of ferromagnetic domains in a CeAlSi crystal. The sample was field-cooled through the transition. (a) shows the in-phase AC susceptometry, (b) shows the DC magnetometry. The light blue areas in the in-phase

*component of AC susceptibility correspond to the stable domains (the white smeared mark in the middle is an artifact due to the saturation of the lock-in amplifier by large signals from point dipoles on the sample surface seen in the magnetometry). The orange arrows at bottom indicate the orientation of external magnetic field relative to the scanning area. The domain walls are lined-up with the crystallographic [100] and [010] directions.*

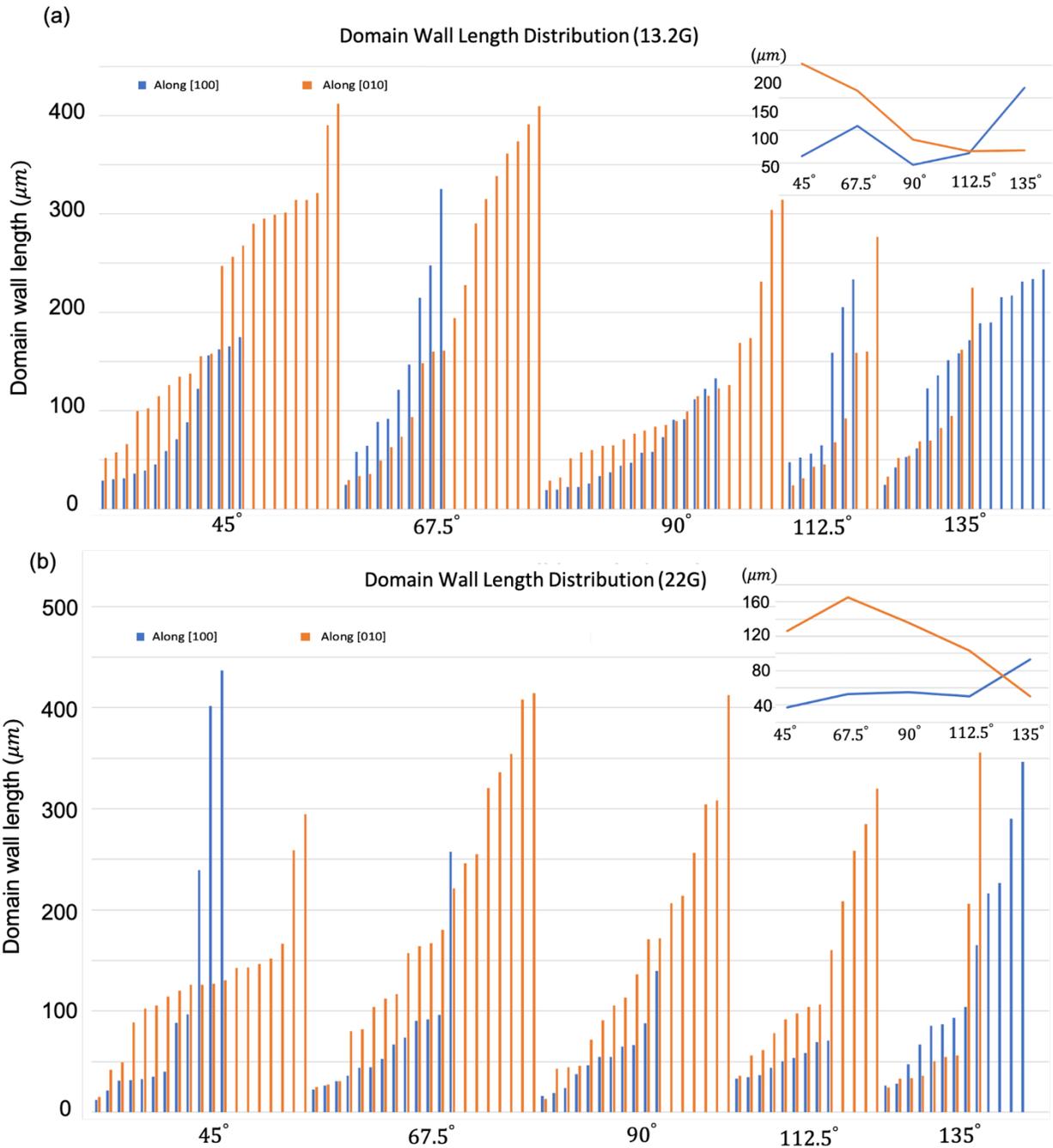

*Figure 3. These histograms show the distribution of domain wall lengths.* *The magnetic field is 13.2G in (a) and 22G in (b). The blue bars represent the length of domain walls that are parallel to [100] (upper left to lower right). The orange bars are the length of domain walls that are oriented in [010]. Insets show the median lengths of domain walls.*